\documentclass[prd,onecolumn,nofootinbib]{revtex4}

\usepackage{amsmath}
\usepackage{graphicx}
\usepackage{amsfonts}
\usepackage{latexsym}
\usepackage{bbold}
\usepackage{wasysym}
\usepackage{calligra}
\usepackage{float}
\usepackage{ulem}
\usepackage{inputenc}
\usepackage{xspace}
\usepackage{url}
\usepackage{epstopdf}
\usepackage{tikz}

\newcommand{\be}{\begin{equation}}
\newcommand{\ee}{\end{equation}}
\newcommand{\beq} {\begin{equation}}
\newcommand{\eeq} {\end{equation}}
\newcommand{\ba}{\begin{eqnarray}}
\newcommand{\ea}{\end{eqnarray}}

\usepackage{microtype}

\begin{document}
	
	\title{Metric-Affine Version of Myrzakulov $F(R,T,Q, {\cal T})$ Gravity and Cosmological Applications}
	
	\author{Damianos Iosifidis$^\ast$, Nurgissa  Myrzakulov$^{\dag,\star}$, Ratbay Myrzakulov$^{\dag,\star}$}
	
	\affiliation{$^\ast$Institute of Theoretical Physics, Department of Physics, Aristotle University of Thessaloniki, 54124 Thessaloniki, Greece. \\
		$^\dag$	Ratbay Myrzakulov  International Centre for Theoretical Physics, Nur-Sultan 010009, Kazakhstan. \\
		$^\star$Eurasian  National University, Nur-Sultan 010008, Kazakhstan.}
	\email{diosifid@auth.gr, nmyrzakulov@gmai.com, rmyrzakulov@gmai.com}
	
	\date{\today}
	
	\begin{abstract}
	 We derive the full set of field equations for the Metric-Affine version of the Myrzakulov gravity model and also extend this family of theories to a broader one. More specifically, we consider theories whose gravitational Lagrangian is given by $F(R,T,Q, {\cal T},{\cal D})$ where $T$, $Q$ are the torsion and non-metricity scalars, ${\cal T}$ is the trace of the energy-momentum tensor and ${\cal D}$ the divergence of the dilation current.   We then consider the linear case of the aforementioned theory and assuming a cosmological setup we obtain the modified Friedmann equations. In addition, focusing on the vanishing non-metricity sector and considering matter coupled to torsion we obtain the complete set of equations describing the cosmological behaviour of this model along with solutions.
		
	\end{abstract}
	
	\maketitle
	
	\allowdisplaybreaks
	
	
	\tableofcontents
	
	
	\section{Introduction}\label{intro}
		
		Even though General Relativity (GR) is undeniably one of the most beautiful and successful theories of physics, recent observational data have challenged its status\cite{will2014confrontation}. Probably the most important observations that cannot be explained within the realm of GR are the early time as well as the late time accelerated expansion of our Universe. This contradiction between theory and observations have lead to the development of a fairly large number of theories alternative to GR which collectively go by the name of Modified Gravity \cite{saridakis2021modified}.   The search of a successful alternative has been proven to  be both fruitful as well as constructive in regards with our understanding of gravity.
		
		Among this plethora of modified gravities let us mention the metric $f(R)$ theories , the Metric-Affine (Palatini) $f(R)$ gravity \cite{sotiriou2008f,iosifidis2019torsion,capozziello2010metric}, the teleparallel $f(T)$ gravities \cite{aldrovandi2012teleparallel,myrzakulov2011accelerating}, the symmetric teleparallel $f(Q)$ \cite{nester1998symmetric,jimenez2018teleparallel}, Scalar-Tensor theories \cite{bartolo1999scalar,charmousis2012general}, etc and also certain extensions of them (see discussion on chapter $IV$). Of course, the kind of modifications one chooses to adopt is highly a matter  of personal taste. In our point of view, interesting and well motivated alternatives are those which extend the underlying geometry of spacetime by allowing a connection more general than the usual Levi-Civita one.  In generic settings, when no a priori restriction is imposed on the connection and the latter is regarded as another  fundamental field on top of the metric the space will be non-Riemannian \cite{eisenhart2012non} and possesses both torsion and non-metricity. These last geometric quantities can then be computed once the affine connection is found. The theories formulated on this non-Riemannian manifold are known as Metric-Affine theories of gravity \cite{hehl1995metric,iosifidis2019metric}.

	In recent years there has been an ever increasing interest  in the Metric-Affine approach \cite{iosifidis2019exactly,iosifidis2019scale,vitagliano2011dynamics,sotiriou2007metric,capozziello2010metric,percacci2020new,Jimenez:2020dpn,BeltranJimenez:2019acz,Olmo:2011uz,aoki2019scalar,Cabral:2020fax,Ariwahjoedi:2020wmo,Yang:2021fjy,helpin2020metric,bahamonde2020new,iosifidis2021parity} and  especially to its cosmological applications \cite{iosifidis2021riemann,iosifidis2020cosmic,Iosifidis:2020gth,Iosifidis:2021iuw,jimenez2016spacetimes,beltran2017modified,kranas2019friedmann,barragan2009bouncing,shimada2019metric,kubota2021cosmological,Mikura:2020qhc,mikura2021minimal}. This interest is possibly due to the fact that the additional affects (compared to GR) that come into play in this framework have a direct geometrical interpretation. That is, the modifications are solely due to spacetime torsion and non-metricity. Furthermore, these geometric notions are excited by matter that has intrinsic structure\cite{hehl1976hypermomentum,babourova1995variational,obukhov1993hyperfluid,Iosifidis:2020gth,Iosifidis:2021nra}. This inner structure-generalized geometry interrelation adds another positive characteristic to the MAG scheme. This is the framework we are going to consider in this study.
	
	In particular, the paper is organized as follows. Firstly, we fix conventions and briefly review some of the basic elements of non-Riemannian geometry and the physics of Metric-Affine gravity. We then consider an extended version of the $F(R,T,Q,{\cal T},{\cal D})$ theory  \cite{myrzakulov2012dark}. To be more specific, working in a Metric-Affine setup, we consider the class of theories with gravitational Lagrangians of the form $F(R,T,Q,{\cal T},{\cal D})$, where ${\cal D}$ is the divergence of the dilation current, the new add-on we are establishing here. Then, we obtain the field equations for this family of theories by varying with respect to the metric and the independent affine connection. Considering a linear function $F$ we then present a cosmological application for this model and finally switching off non-metricity and considering a scalar field coupled to torsion we obtain the modified Friedmann equations and also provide solutions for this simple case.

	\section{Conventions/Notation}
Let us now briefly go over the basic geometric as well as physical setup we are going to use and also	fix notation. We consider a $4$-dim non-Riemannian manifold endowed with  a metric and an affine connection $({\cal M}, g,  \nabla)$. Our definition for the covariant derivative, of a vector say, will be
	\beq
\nabla_{\alpha}u^{\lambda}=\partial_{\alpha}u^{\lambda}+\Gamma^{\lambda}_{\;\;\;\beta\alpha}u^{\beta}
\eeq
We also define the (Cartan)
torsion  tensor by
\beq
S_{\mu\nu}^{\;\;\;\;\lambda}:=\Gamma^{\lambda}_{\;\;\;[\mu\nu]}
\eeq
and the non-metricity tensor as
\beq
Q_{\alpha\mu\nu}:=-\nabla_{\alpha}g_{\alpha\beta}
\eeq
Contracting these with the metric tensor we obtain the associated torsion and non-metricity vectors
\beq
S_{\mu}:=S_{\mu\nu}^{\;\;\;\;\nu}
\eeq
\beq
 Q_{\mu}:=Q_{\mu\alpha\beta}g^{\alpha\beta}\;\;,\;\; q_{\mu}:=Q_{\alpha\beta\mu}g^{\alpha\beta}
\eeq
respectively. 
In addition, since we are in four dimensions, we can also form the torsion pseudo-vector according to
\beq
t^{\mu}:=\varepsilon^{\mu\alpha\beta\gamma}S_{\alpha\beta\gamma}
\eeq
Given the above definitions for torsion and non-metricity one can easily show (see for instance \cite{iosifidis2019metric}) the affine connection decomposition\footnote{From here onwards we shall use the tilde notation in order to denote Riemannian objects, that is objects computed with respect to the Levi-Civita connection $\tilde{\Gamma}^{\lambda}_{\;\;\;\mu\nu}$. }
\begin{gather}
\Gamma^{\lambda}_{\;\;\;\mu\nu}=N^{\lambda}_{\;\;\;\;\mu\nu}+\tilde{\Gamma}^{\lambda}_{\;\;\;\mu\nu}=
\frac{1}{2}g^{\alpha\lambda}(Q_{\mu\nu\alpha}+Q_{\nu\alpha\mu}-Q_{\alpha\mu\nu}) -g^{\alpha\lambda}(S_{\alpha\mu\nu}+S_{\alpha\nu\mu}-S_{\mu\nu\alpha}) \label{N}
\end{gather}
where $N^{\lambda}_{\;\;\mu\nu}$ is known as the distortion tensor.
Continuing we define the curvature tensor as usual
	\beq
R^{\mu}_{\;\;\;\nu\alpha\beta}:= 2\partial_{[\alpha}\Gamma^{\mu}_{\;\;\;|\nu|\beta]}+2\Gamma^{\mu}_{\;\;\;\rho[\alpha}\Gamma^{\rho}_{\;\;\;|\nu|\beta]} \label{R}
\eeq
and by a double contraction of the latter, we get the Ricci scalar
\beq
R:=R^{\mu}_{\;\;\;\nu\mu\beta}g^{\nu\beta}
\eeq
Then by using decomposition ($\ref{N}$) we obtain the post Riemannian expansion for the Ricci scalar \cite{iosifidis2019metric}
	\beq
R=\tilde{R}+T+Q +2 Q_{\alpha\mu\nu}S^{\alpha\mu\nu}+2 S_{\mu}(q^{\mu}-Q^{\mu}) +\tilde{\nabla}_{\mu}(q^{\mu}-Q^{\mu}-4S^{\mu})
\eeq
where $\tilde{R}$ is the Riemannian Ricci tensor (i.e. computed with respect to the Levi-Civita connection) and we have also defined the torsion and non-metricity scalars as
	\beq
T:= S_{\mu\nu\alpha}S^{\mu\nu\alpha}-2S_{\mu\nu\alpha}S^{\alpha\mu\nu}-4S_{\mu}S^{\mu} \label{Tsc}
\eeq
and 
	\beq
Q:= \frac{1}{4}Q_{\alpha\mu\nu}Q^{\alpha\mu\nu}-\frac{1}{2}Q_{\alpha\mu\nu}Q^{\mu\nu\alpha}    -\frac{1}{4}Q_{\mu}Q^{\mu}+\frac{1}{2}Q_{\mu}q^{\mu} \label{Qsc}
\eeq
respectively. Note that with the introduction of the superpotentials\footnote{Here we are using the conventions of \cite{iosifidis2019scale}.}
\beq
\Omega^{\alpha\mu\nu} := \frac{1}{4} Q^{\alpha\mu\nu}-\frac{1}{2} Q^{\mu\nu\alpha}-\frac{1}{4} g^{\mu\nu}Q^{\alpha}+\frac{1}{2}g^{\alpha\mu}Q^{\nu}
\eeq
\beq
\Sigma^{\alpha\mu\nu} :=S^{\alpha\mu\nu}-2S^{\mu\nu\alpha}-4g^{\mu\nu}S^{\alpha}
\eeq 
these can more compactly be expressed as
\beq
T=S_{\alpha\mu\nu}\Sigma^{\alpha\mu\nu}
\eeq
	\beq
Q=Q_{\alpha\mu\nu}\Omega^{\alpha\mu\nu}
\eeq

 Equation ($\ref{R}$) is of key importance in teleparallel formulations. For instance by imposing vanishing curvature (which also implies $R=0$) and metric compatibility ($Q_{\alpha\mu\nu}=0$)
one obtains from $(\ref{N})$,
	\beq
\tilde{R}=-T +4 \tilde{\nabla}_{\mu}S^{\mu}
\eeq
which is the basis of the metric teleparallel formulation. In a similar manner the symmetric teleparallel (vanishing curvature and torsion) and also the generalized teleparalelism (only vanishing curvature) are obtained \cite{jimenez2019general}.

Let us now turn our attention to the matter content. In Metric-Affine gravity apart from the  energy momentum tensor, which we define as usual
\beq
T_{\mu\nu}:=-\frac{2}{\sqrt{-g}}\frac{\delta(\sqrt{-g} \mathcal{L}_{M})}{\delta g^{\mu\nu}}
\eeq
one also has to vary the matter part with respect to the affine-connection.  This new object, which is defined  by
\beq
\Delta_{\lambda}^{\;\;\;\mu\nu}:= -\frac{2}{\sqrt{-g}}\frac{\delta ( \sqrt{-g} \mathcal{L}_{M})}{\delta \Gamma^{\lambda}_{\;\;\;\mu\nu}}
\eeq
is called hypermomentum \cite{hehl1976hypermomentum} and encodes the microscopic characteristics of matter such as spin, dilation and shear. In the same way that the energy momentum tensor sources spacetime curvature by means of the metric field equations, the hypermomentum is the source of spacetime torsion and non-metricity (through the connection field equations). Note that these energy related tensors are not quite independent and are subject to the conservation law
\beq
\sqrt{-g}(2 \tilde{\nabla}_{\mu}T^{\mu}_{\;\;\alpha}-\Delta^{\lambda\mu\nu}R_{\lambda\mu\nu\alpha})+\hat{\nabla}_{\mu}\hat{\nabla}_{\nu}(\sqrt{-g}\Delta_{\alpha}^{\;\;\mu\nu})+2S_{\mu\alpha}^{\;\;\;\;\lambda}\hat{\nabla}_{\nu}(\sqrt{-g}\Delta_{\lambda}^{\;\;\;\mu\nu})=0\;, \;\; \hat{\nabla}_{\mu}:=2S_{\mu}-\nabla_{\mu} \label{ccc}
\eeq
which comes from the diffeomorphism invariance of the matter sector of the action (see \cite{Iosifidis:2020gth}). In the above discussion we have briefly developed the geometric and physical setup needed for the rest of our study.  Let us focus on the cosmological aspects of theories with torsion and non-metricity (i.e non-Riemannian extensions). 

		\section{Cosmology with Torsion and Non-metricity}
Let us consider a homogeneous flat FLRW Cosmology, with the usual Robertson-Walker line element
	\beq
	ds^{2}=-dt^{2}+a^{2}(t)\delta_{ij}dx^{i}dx^{j} \label{metric}
	\eeq
	where 	$i,j=1,2,3$ and $a(t)$ is as usual the scale factor of the universe. As usual  the Hubble parameter is defined as $H:=\dot{a}/a$. Now, let $u^{\mu}$ be the normalized $4$-velocity field  and
	\beq
	h_{\mu\nu}:=g_{\mu\nu}+u_{\mu}u_{\nu}
	\eeq
be the projection tensor	projecting objects on the space orthogonal to $u^{\mu}$. 
 The affine connection of the non-Riemannian FLRW spacetime reads \cite{Iosifidis:2020gth}
\beq
\Gamma^{\lambda}_{\;\;\;\mu\nu}=\widetilde{\Gamma}^{\lambda}_{\;\;\;\mu\nu}+	X(t)u^{\lambda}h_{\mu\nu}+Y(t)u_{\mu}h^{\lambda}_{\;\;\nu}+Z(t)u_{\nu}h^{\lambda}_{\;\;\mu}
+V(t)u^{\lambda}u_{\mu}u_{\nu} +\epsilon^{\lambda}_{\;\;\mu\nu\rho}u^{\rho}W(t)\delta_{n,4} \label{connect}
\eeq
where the non-vanishing components of the Levi-Civita connection are in this case
\beq
\tilde{\Gamma}^{0}_{\;\; ij}=\tilde{\Gamma}^{0}_{\;\; ji}=\dot{a}a \delta_{ij}=H g_{ij}
\;, \;\;
\tilde{\Gamma}^{i}_{\;\; j0}=\tilde{\Gamma}^{i}_{\;\; 0j}=\frac{\dot{a}}{a}\delta^{i}_{\;\;j}=H\delta^{i}_{\;j}
\eeq
Continuing with the rest of the geometric objects, in this highly symmetric spacetime, the torsion and non-metricity tensors take the forms \cite{Iosifidis:2020gth}  
	\beq
	S_{\mu\nu\alpha}^{(n)}=2u_{[\mu}h_{\nu]\alpha}\Phi(t)+\epsilon_{\mu\nu\alpha\rho}u^{\rho}P(t) \label{Scosm}
	\eeq
	\beq
	Q_{\alpha\mu\nu}=A(t)u_{\alpha}h_{\mu\nu}+B(t) h_{\alpha(\mu}u_{\nu)}+C(t)u_{\alpha}u_{\mu}u_{\nu}  \label{Qcosm}
	\eeq
	respectively. The five functions $\Phi,P,A,B,C$ describe  the non-Riemannian Cosmological effects. These, along with the scale factor, give the cosmic evolution of non-Riemannian geometries. Let us note that using the relations of the torsion and non-metricity tensors with the distortion tensor it is trivial to show that
	 the functions $X(t),Y(t),Z(t),V(t),W(t)$ are linearly related to $\Phi(t),P(t),A(t),B(t),C(t)$ as \cite{Iosifidis:2020gth}
	\beq
	2(X+Y)=B \;, \;\; 2Z=A\;, \;\; 2V=C \;, \;\; 2\Phi =Y-Z\;, \;\; P = W	
	\eeq
	or inverting them
	\beq
	W=P \;, \;\; V=C/2 \;, \;\; Z=A/2	
	\eeq
	\beq
	Y=2\Phi +\frac{A}{2}	\;\;, \;\;\;
	X=\frac{B}{2}- 2 \Phi -\frac{A}{2} \label{XY}
	\eeq
Now, using the definitions ($\ref{Tsc}$) and $(\ref{Qsc})$ for the torsion and non-metricity scalars and the above cosmological forms for torsion and non-metricity we find for the former
	\beq
T	=24\Phi^{2}-6P^{2}
\eeq
\begin{gather}
Q=\frac{3}{4}\left[ 2A^{2}+B(C-A) \right]
\end{gather}
respectively. These are the expressions for the torsion and non-metricity scalars in a homogeneous cosmological setup when no teleparallelism is imposed.

	Finally, using the post Riemannian decomposition of the Ricci scalar and the above forms of the torsion and non-metricity scalars we find
	\begin{gather}
	R=\tilde{R}+6\left[ \frac{1}{4}A^{2}+4 \Phi^{2}+\Phi (2A-B) \right]+\frac{3}{4} B (C-A) - 6 P^{2}\nonumber \\
	+3\frac{1}{\sqrt{-g}}\partial_{\mu} \left[ \sqrt{-g} u^{\mu}\left( \frac{B}{2}-A-4 \Phi \right)    \right] \label{Rpost}
	\end{gather}
	where
	\beq
\tilde{R}=6 \left[ \frac{\ddot{a}}{a}+\left(\frac{\dot{a}}{a}\right)^{2} \right]
\eeq
	is the usual Riemannian part. The last decomposition will be very useful in our subsequent discussion.

	\section{MG-VIII model and Extension: The $F(R,T,Q,{\cal T},{\cal D})$ Theories }
	In this paper, we are going to study  the Myrzakulov gravity \cite{myrzakulov2012dark} VIII (MG-VIII)\footnote{See also \cite{anagnostopoulos2021observational,Saridakis:2019qwt} for some observational implications of this theory.}. Its action is given by \cite{myrzakulov2012dark}
	\begin{equation}
	S[g,\Gamma, \phi]=S_{g}+S_{m}=\frac{1}{2\kappa}\int \sqrt{-g}d^{4}x \left[F(R,T,Q,{\cal T})+2\kappa \mathcal{L}_{m}\right], \label{2.1}
	\end{equation}
	where  $R$ stands for the Ricci scalar (curvature scalar), $T$ is the torsion scalar, $Q$ is the nonmetricity scalar and  ${\cal T}$ is trace of the energy-momentum tensor of matter Lagrangian $L_{m}$. The MG-VIII  can be seen as some kind of unification of $F(R), F(T), F(Q)$  or $F(R,{\cal T}), F(T, {\cal T}), F(Q, {\cal T})$ theories (see \cite{Harko:2011kv,harko2014f,xu2019f} respectively). For instance, if one imposes flatness (i.e. $R^{\lambda}_{\;\;\alpha\mu\nu} \equiv0$) and metric compatibility ($Q_{\alpha\mu\nu}\equiv 0$) arrives at the $f(T)$ gravity \cite{myrzakulov2011accelerating,krvsvsak2016covariant}. Demanding flatness and a  torsionless connection we get symmetric teleparallel $f(Q)$ gravity \cite{nester1998symmetric,jimenez2018teleparallel}. More generally, imposing only teleparallelism we arrive at the recently developed generalized teleparallel scheme  of f(G) \cite{jimenez2019general,beltran2021accidental} theories. If no restriction on the connection is assumed then $(\ref{2.1})$ serves as a specific generalization of metric-affine $f(R)$ gravity where where the energy momentum trace $\cal T$ and certain quadratic combinations of torsion and non-metricity are added as well. In fact in this generalized metric-affine setup one could consider also the presence of the hypermomentum analogue of the (metrical) energy momentum trace. Giving it a little thought we observe that similar to the trace $\cal T$ is the divergence of the dilation current as they appear in the trace of the canonical\footnote{Here $t=t^{\mu\nu}g_{\mu\nu}$ is the trace of the canonical energy momentum tensor $t^{\mu\nu}$.} energy momentum tensor (see for instance \cite{Iosifidis:2020gth})
		\beq
	t={\cal T}+\frac{1}{2 \sqrt{-g}}\partial_{\nu}(\sqrt{-g}\Delta^{\nu}) \;\;, \;\;\; \Delta^{\nu}:=\Delta_{\mu}^{\;\;\mu\nu}
	\eeq
	In this sense ${\cal T}$ and  the divergence of $\Delta^{\nu}$ are placed on equal footing as it is obvious from the above equation.
	Therefore, the scalar obtained by the divergence of the dilation current
	\beq
{\cal D}=\frac{1}{ \sqrt{-g}}\partial_{\nu}(\sqrt{-g}\Delta^{\nu}) 	
	\eeq
	would be trace analogue for the hypermomentum. With this inclusion we may generalize the class of theories (\ref{2.1}) to
	\begin{equation}
	S[g,\Gamma, \phi]=S_{g}+S_{m}=\frac{1}{2\kappa}\int \sqrt{-g}d^{4}x \left[F(R,T,Q,{\cal T}, {\cal D})+2\kappa \mathcal{L}_{m}\right], \label{div}
	\end{equation}

	The field equations of the family of theories given by the the above action read as follows:
	
	\textbf{g-Variation:}
	\begin{gather}
	-\frac{1}{2}g_{\mu\nu}F+F_{R}R_{(\mu\nu)}+F_{T}\Big(2S_{\nu\alpha\beta}S_{\mu}^{\;\;\;\alpha\beta}-S_{\alpha\beta\mu}S^{\alpha\beta}_{\;\;\;\;\nu}+2S_{\nu\alpha\beta}S_{\mu}^{\;\;\;\beta\alpha}-4S_{\mu}S_{\nu} \Big)+F_{Q}L_{(\mu\nu)}\nonumber \\
	+\hat{\nabla}_{\lambda}(F_{Q}J^{\lambda}_{\;\;\;(\mu\nu)})+g_{\mu\nu}\hat{\nabla}_{\lambda}(F_{Q}\zeta^{\lambda})+F_{\cal T}(\Theta_{\mu\nu}+T_{\mu\nu})+F_{D}M_{\mu\nu}=\kappa T_{\mu\nu}
	\end{gather}
	where 
	\beq
	\hat{\nabla}_{\lambda}:= \frac{1}{\sqrt{-g}}(2S_{\lambda}-\nabla_{\lambda})
	\eeq
	
	\beq
	\Omega^{\alpha\mu\nu} = \frac{1}{4}Q^{\alpha\mu\nu}-\frac{1}{2} Q^{\mu\nu\alpha}-\frac{1}{4} g^{\mu\nu}Q^{\alpha}+\frac{1}{2}g^{\alpha\mu}Q^{\nu}
	\eeq
	\begin{gather}
	4 L_{\mu\nu}=(Q_{\mu\alpha\beta}-2 Q_{\alpha\beta\mu})Q_{\nu}^{\;\;\;\alpha\beta}+(Q_{\mu}+2q_{\mu})Q_{\nu}
	+(2Q_{\mu\nu\alpha}-Q_{\alpha\mu\nu})Q^{\alpha})\nonumber \\-4 \Omega^{\alpha\beta}_{\;\;\;\;\nu}Q_{\alpha\beta\mu}-4 \Omega_{\alpha\mu\beta}Q^{\alpha\beta}_{\;\;\;\;\nu}
	\end{gather}
	\beq
	\Theta_{\mu\nu}:=g^{\alpha\beta}\frac{\delta T_{\alpha\beta}}{\delta g^{\mu\nu}}
	\eeq
	\beq
	M_{\mu\nu}:=\frac{\delta D}{\delta g^{\mu\nu}}
	\eeq
	and we have also defined the densities
	\beq
	J^{\lambda}_{\;\;\;\mu\nu} := \sqrt{-g}\Big( \frac{1}{4} Q^{\lambda}_{\;\;\;\mu\nu}-\frac{1}{2}Q_{\mu\nu}^{\;\;\;\;\lambda}+\Omega^{\lambda}_{\;\;\;\mu\nu}\Big)
	\eeq
	\beq
	\zeta^{\lambda}=\sqrt{-g}\Big(-\frac{1}{4}Q^{\lambda}+\frac{1}{2}q^{\lambda}\Big)
	\eeq
	
	\textbf{$\Gamma$-Variation:}
	\begin{gather}
	P_{\lambda}^{\;\;\mu\nu}(F_{R})+2 F_{T}\Big( S^{\mu\nu}_{\;\;\;\;\lambda}-2 S_{\lambda}^{\;\;\;[\mu\nu]}-4 S^{[\mu}\delta^{\nu]}_{\lambda}\Big)-M_{\lambda}^{\;\;\mu\nu\alpha}\partial_{\alpha}F_{D} \nonumber \\
	+F_{Q}\Big( 2 Q^{[\nu\mu]}_{\;\;\;\;\lambda}-Q_{\lambda}^{\;\;\mu\nu}+(q^{\nu}-Q^{\nu})\delta^{\mu}_{\lambda}+Q_{\lambda}g^{\mu\nu}+\frac{1}{2}Q^{\mu}\delta^{\nu}_{\lambda} \Big)=F_{\cal T}\Theta_{\lambda}^{\;\;\mu\nu}+\kappa \Delta_{\lambda}^{\;\;\mu\nu}
	\end{gather}
	where
	\begin{gather}
	P_{\lambda}^{\;\;\;\mu\nu}(F_{R}) = -\frac{\nabla_{\lambda}(\sqrt{-g}F_{R}g^{\mu\nu})}{\sqrt{-g}}+\frac{\nabla_{\alpha}(\sqrt{-g}F_{R}g^{\mu\alpha}\delta_{\lambda}^{\nu})}{\sqrt{-g}}+ \\ \nonumber
	2 F_{R}(S_{\lambda}g^{\mu\nu}-S^{\mu}\delta_{\lambda}^{\nu}-  S_{\lambda}^{\;\;\;\mu\nu}) 
	\end{gather}
	is the modified Palatini tensor and 
	\beq
	\Theta_{\lambda}^{\;\;\mu\nu}:=-\frac{\delta \cal T}{\delta \Gamma^{\lambda}_{\;\;\;\mu\nu}} \;\;, \;\; M_{\lambda}^{\;\;\mu\nu\alpha}:=\frac{\delta \Delta^{\alpha}}{\delta \Gamma^{\lambda}_{\;\;\;\mu\nu}} 
	\eeq
	Note: If matter does not couple to the connection (e.g. classical perfect fluid with no inner structure) we have that $\Theta_{\lambda}^{\;\;\mu\nu}=0$ as well as $ \Delta_{\lambda}^{\;\;\mu\nu}=0$ and $M_{\lambda}^{\;\;\mu\nu\alpha}$. The above set of field equations constitutes an extended (with the divergence of dilation included) Metric-Affine version of the Myrzakulov gravities \cite{myrzakulov2012dark}. Here we derived the field equations with no restriction on the connection and also for the extended case $F(R,T,Q,{\cal T}, D)$. In the sequel we shall analyse further the linear case $F=R+\beta T+\gamma Q+\mu {\cal T}+\nu {\cal D}$ and also touch upon cosmological applications.

	\section{Cosmological Applications}

	\subsection{The Cosmology of $F=R+\beta T+\gamma Q+\mu {\cal T}+\nu {\cal D}$ Theory}
	Let us now analyse in more detail the linear case $$F=R+\beta T+\gamma Q+\mu {\cal T}+\nu {\cal D}$$ and also obtain the associated cosmological equations. To start with let us note that since $\sqrt{-g}{\cal D}$ is a total divergence, the dilation current does not contribute to the field equations when included linearly. Therefore we can safely set $\nu=0$ for the rest of our discussion. In addition, in this linear case the  metric field equations take the form  
		\begin{gather}
	-\frac{1}{2}g_{\mu\nu}F+R_{(\mu\nu)}+\beta\Big(2S_{\nu\alpha\beta}S_{\mu}^{\;\;\;\alpha\beta}-S_{\alpha\beta\mu}S^{\alpha\beta}_{\;\;\;\;\nu}+2S_{\nu\alpha\beta}S_{\mu}^{\;\;\;\beta\alpha}-4S_{\mu}S_{\nu} \Big)+\gamma L_{(\mu\nu)}\nonumber \\
	+\hat{\nabla}_{\lambda}(\gamma J^{\lambda}_{\;\;\;(\mu\nu)})+g_{\mu\nu}\hat{\nabla}_{\lambda}(\gamma \zeta^{\lambda})+\mu(\Theta_{\mu\nu}+T_{\mu\nu})=\kappa T_{\mu\nu} \label{Pt}
	\end{gather}
	Taking the trace of the last equation, using the post Riemannian expansion ($\ref{Rpost}$)  and also employing ($\ref{Scosm}$) along with ($\ref{Qcosm}$) and after some long calculations we finally arrive at
	\begin{gather}
	\frac{\ddot{a}}{a}+\left(\frac{\dot{a}}{a}\right)^{2}+(1+\beta)( 4 \Phi^{2}-P^{2})+\frac{1}{8}\Big( 2 A^{2}+B(C-A) \Big) +\Phi(2 A-B)+\dot{f}+3 H f=-\mu (\Theta+{\cal T})+\kappa {\cal T}
	\end{gather}
	where 
	\beq
f:=\frac{1}{2}\left[ (1-\gamma)\Big( \frac{B}{2}-A \Big)	-4 \Phi \right] \;, \;\; \Theta:=\Theta_{\mu\nu}g^{\mu\nu}
	\eeq
	which is a variant of the modified Friedmann equation. As for the second Friedmann (acceleration) equation, its general form was derived in \cite{iosifidis2020cosmic} for general non-Riemannian cosmological setups. It reads
	\begin{gather}
	\frac{\ddot{a}}{a}=-\frac{1}{3}R_{\mu\nu}u^{\mu}u^{\nu}+2\left( \frac{\dot{a}}{a} \right)\Phi +2\dot{\Phi} 
	+\left( \frac{\dot{a}}{a} \right)\left(A+\frac{C}{2}\right)+\frac{\dot{A}}{2}-\frac{A^{2}}{2}-\frac{1}{2}AC  
	-2A\Phi-2C \Phi\label{accel}
	\end{gather}
	One could then proceed by contracting  $(\ref{Pt})$ with $u^{\mu}u^{\nu}$ in order to eliminate the first term ($R_{\mu\nu}u^{\mu}u^{\nu}$) and express everything in terms of the scale factor and the torsion and non-metricity variables. This results in a fairly complicated expression which we refrain from presenting it here since it goes beyond the scope of the present exposure. As a final note let us mention that in order to analyse in depth the above cosmological model one should consider an appropriate form of matter for which both the metrical energy momentum and hypermomentum tensors must respect the cosmological principle. The fluid with such characteristics was constructed in \cite{Iosifidis:2020gth} (see also for a generalized version \cite{Iosifidis:2021nra}) and goes by the name Perfect Cosmological Hyperfluid. The hypermomentum part of this fluid will then source the torsion and non-metricity variables $\Phi, P,A,...$ etc. by virtue of the connection field equations. We note that scalar fields coupled to the connection belong (are certain subcases) to the aforementioned fluid  description. For the sake of illustration, below we present such an example with a scalar field non-minimally coupled to the connection in the case of vanishing non-metricity and also study some of the cosmological implications of this theory.

\subsection{Scalar Field Coupled to Torsion}
We shall now focus on the vanishing non-metricity sector and also set $\gamma=0$ , that is we will concentrate on the case $F=R+ \beta T$.  As for the matter part let us consider a scalar field. In the usual (i.e. purely Riemannian) case one would have the usual Lagrangian
\beq
	{\cal L}^{(0)}_{m}=-\frac{1}{2}g^{\mu\nu}\nabla_{\mu}\phi\nabla_{\nu}\phi-V(\phi), \label{phi}
\eeq
for the scalar field $\phi$. However, in the presence of torsion nothing prevent us to consider direct couplings of the scalar field with torsion. The most straightforward form of such a coupling is a torsion vector-scalar field derivative interaction of the form $\lambda_{0}S^{\mu}\nabla_{\mu}\phi$, where $\lambda_{0}$ is the coupling constant measuring the strength of the interaction. Including this term, our full matter Lagrangian now reads
\beq
	{\cal L}_{m}=-\frac{1}{2}g^{\mu\nu}\nabla_{\mu}\phi\nabla_{\nu}\phi-V(\phi)+\lambda_{0}S^{\mu}\nabla_{\mu}\phi \label{phi2}
\eeq
Then, substituting this into $(\ref{2.1})$ and varying the latter with respect to the scalar field, we obtain
\beq
\frac{1}{\sqrt{-g}}\partial_{\mu}\Big[ \sqrt{-g}(\partial^{\mu}\phi-\lambda_{0}S^{\mu})\Big]=\frac{\partial V}{\partial \phi} \label{phieom}
\eeq
which is the evolution equation for the scalar field under the influence of torsion. 
In addition, the very presence of the interaction term $\lambda_{0}S^{\mu}\nabla_{\mu}\phi$ produces a non-vanishing hypermomentum which is trivially computed to be
\beq
\Delta_{\lambda}^{\;\;\mu\nu}=2 \lambda_{0}\delta^{[\mu}_{\lambda}\nabla^{\nu]}\phi
\eeq
 With this result, starting from the connection field equations () which in our case read
\begin{gather}
P_{\lambda}^{\;\;\mu\nu}+2 \beta\Big( S^{\mu\nu}_{\;\;\;\;\lambda}-2 S_{\lambda}^{\;\;\;[\mu\nu]}-4 S^{[\mu}\delta^{\nu]}_{\lambda}\Big)=\kappa \Delta_{\lambda}^{\;\;\mu\nu} \label{PS}
\end{gather}
and contracting in $\mu=\lambda$ we find
	\beq
	S^{\mu}=\frac{3 \kappa \lambda_{0}}{8 \beta}\partial^{\mu}\phi \label{sphi}
	\eeq
	that is the presence of scalar field produces spacetime torsion\footnote{Of course this is so because of the connection coupling which yields a non-vanishing hypermomentum. If no such coupling is included the scalar field cannot either feel or produce torsion.}.
	In addition, contracting ($\ref{PS}$) with $ \varepsilon^{\lambda}_{\;\;\mu\nu\alpha}$ it follows that
	\beq
	t_{\alpha}=0 \label{t}
	\eeq
	Note that we can now plug back to (\ref{phi2}) the above form of the torsion tensor to end up with
	\beq
		{\cal L}_{m}=-\frac{1}{2}\left(1-\frac{3 \kappa \lambda_{0}^{2}}{4 \beta} \right)g^{\mu\nu}\nabla_{\mu}\phi\nabla_{\nu}\phi-V(\phi) \label{phi3}
	\eeq
	Interestingly, from the last equation we conclude that the scalar-torsion interaction changes the factor of the kinetic term for the scalar field. We also see that the is a crucial value for the coupling $|\lambda_{0}|=2 \sqrt{\frac{\beta}{3 \kappa}}$ above which the kinetic term changes sign and for exactly this value vanishes identically. Since this last case would require severe fine tuning we shall disregard it and we shall also assume that $\lambda_{0}$ is under this bound so that the kinetic term keeps its original sign.

	Up to this point,	the above considerations were general. Let us now focus on the homogeneous FLRW cosmology of this theory. In this case, equation ($\ref{t}$) implies that $P=0$	
	and as a result, upon using $(\ref{sphi})$ the full torsion tensor is given by	
	\beq
	S_{\mu\nu\alpha}=2u_{[\mu}h_{\nu]\alpha}\Phi(t)  \;\;, \;\; \Phi=-\frac{\kappa \lambda_{0}}{8 \beta}\dot{\phi} \label{ss}
	\eeq
	In the case of a free scalar field (i.e $V(\phi)=0$) inserting ($\ref{sphi}$) into ($\ref{phieom}$) we obtain
	\beq
\left( 1-\frac{3 \kappa \lambda_{0}^{2}}{8 \beta}\right)	\partial_{\mu}\Big[ \sqrt{-g}\partial^{\mu}\phi\Big]=0
	\eeq
	which for $|\lambda_{0}|\neq 2 \sqrt{\frac{\beta}{3 \kappa}}$ implies that
	\beq
\dot{\phi}=\frac{c_{0}}{a^{3}}	\label{dotphi}
	\eeq
	On the other hand, the metric field equations in this case read
		\begin{gather}
	-\frac{1}{2}g_{\mu\nu}F+R_{(\mu\nu)}+\beta \Big(2S_{\nu\alpha\beta}S_{\mu}^{\;\;\;\alpha\beta}-S_{\alpha\beta\mu}S^{\alpha\beta}_{\;\;\;\;\nu}+2S_{\nu\alpha\beta}S_{\mu}^{\;\;\;\beta\alpha}-4S_{\mu}S_{\nu} \Big)=\kappa T_{\mu\nu}
	\end{gather}
	and by taking the trace, using the same procedure we outlined previously,  we finally obtain 
	\beq
\frac{\ddot{a}}{a}+\left(\frac{\dot{a}}{a}\right)^{2} =\left[ -\frac{\kappa}{6}+(1-\beta)\left(\frac{\kappa \lambda_{0}}{4 \beta}\right)^{2}\right]\dot{\phi}^{2} \label{F1}
	\eeq
which is again a variant of the modified Friedmann equation. Let us now derive the acceleration equation for this case. First, we  contract the above field equations with $u^{\mu}u^{\nu}$ to obtain
\beq
R_{\mu\nu}u^{\mu}u^{\nu}=24 \beta \Phi^{2}+\frac{\kappa}{2}(\rho+3 p)
\eeq
which when substituted in $(\ref{accel})$	 for vanishing non-metricity and the given scalar matter results in the acceleration equation
\beq
\frac{\ddot{a}}{a}=- 8 \beta \Phi^{2}-\frac{\kappa}{6}(\rho+3 p)+2 H \Phi+2 \dot{\Phi}
\eeq
where $\rho$ and $p$ are the density and pressure associated to the scalar field Lagrangian $(\ref{phi3})$. It is interesting to note that the first term on the right hand side of the acceleration equation has a fixed sign depending on the value of $\beta$. Intriguingly,  for $\beta<0$ the contribution from this term has always a fixed positive sign producing an accelerated expansion regardless of the sign of $\Phi$ (or equivalently $\dot{\phi}$). As for the last two terms, combining $(\ref{ss}b)$ and $(\ref{dotphi})$ we observe that $\dot{\Phi}=-3 H \Phi$ which when substituted to the above acceleration equation yields
\beq
\frac{\ddot{a}}{a}=- 8 \beta \Phi^{2}-\frac{\kappa}{6}(\rho+3 p)+\frac{4}{3}\dot{\Phi}
\eeq
We can conclude therefore that the last term aids to to acceleration when $\dot{\Phi}>0$ and slows it down whenever $\dot{\Phi}<0$. From the above analysis we see that the  non-Riemannian degrees of freedom play a crucial role on the cosmological evolution providing new interesting phenomena. Now using the latter form of the acceleration equation we can obtain the first Friedmann equation from ($\ref{F1}$) by eliminating the double derivative of the scale factor. For the simple case $V(\phi)=0$
we find 
\beq
\left(\frac{\dot{a}}{a}\right)^{2}=\left[ \frac{\kappa}{6}+(1+\beta)\left(\frac{\kappa \lambda_{0}}{4 \beta}\right)^{2}\right] \dot{\phi}^{2}-\frac{4}{3}\dot{\Phi}
\eeq
	as the modified first Friedmann equation. Note that on substituting ($\ref{sphi}b$) in the above and completing the square in the resulting expression we easily find the power-law solution
	\beq
a(t)\propto t^{1/3}	
	\eeq 
	which is the stiff matter solution. We see that in the simplified case of a zero potential  for the scalar we arrive at a known solution. However, we should remark that the situation changes drastically when one considers a non-vanishing potential. Note also that the torsion tensor in this case goes like $1/t$ and therefore its effect diminishes with time. 
	
	 Needless to say that when non-metricity is also included one gets  more complicated expressions with a much richer phenomenology. It would be quite interesting to see exactly to what degree  the simultaneous presence of torsion and non-metricity alters the cosmological evolution in such models.
	This would be the theme of a separate work.
	
	\section{Conclusions}
	
	By working in a Metric-Affine approach (i.e. considering the metric and the connection as independent variables) we have considered a generalized version of the theory proposed in \cite{myrzakulov2012dark}. In particular,  we derived the full set of field equations of the class of theories whose gravitational part of the Lagrangian is given by $F(R,T,Q,{\cal T}, {\cal D})$, where $T$, $Q$ are the torsion and non-metricity scalars, ${\cal T}$ is the trace of the energy-momentum tensor and ${\cal D}$ is the divergence of the dilation current (one of the hypermomentum sources). The family of theories contained in our Lagrangian is fairly large since all, metric and Palatini $f(R)$ theories, teleparallel $f(T)$, symmetric teleparallel $f(Q)$ or even generalized teleparallel $f(G)$ and generalizations of them such as $f(R, {\cal T})$, $f(T,{\cal T} )$, $f(Q,{\cal T} )$ can be seen as special cases of our theory.

	 Our contribution was two-fold. Firstly, we generalized the family of theories to those including also the  divergence of the dilation current (which is the analogue of the energy-momentum trace for hypermomentum). Furthermore, as already mentioned above, we worked in a Metric-Affine framework, considering an independent affine connection as a fundamental variable along with the metric. This allows one not only to study the aforementioned theories (by restricting the connection one way or another), but also to analyse them in this general Metric-Affine scheme.
	Having derived the complete set of Metric-Affine $F(R,T,Q,{\cal T}, {\cal D})$ theories we then concentrated our attention on the linear case $F=R+\beta T+\gamma Q+\mu {\cal T}+\nu {\cal D}$ and obtained a variant version of the modified Friedmann equation. Finally, we focused on the vanishing non-metricity sector and also considered a scalar field coupled to torsion as our matter sector. In this case we derived both the first and second (acceleration) Friedmann equations and examined under what circumstances the presence of torsion can have an accelerating affect on the cosmological evolution. For this simple case we were also able to provide an exact power-law solution for the scale factor.
	
	In closing let us note some further applications and additional developments of our study here. Firstly, it would be interesting to study in more detail the linear case especially in regard with its cosmological implications in the presence of the cosmological hyperfluid \cite{Iosifidis:2020gth,Iosifidis:2021nra}. In addition, as we have already mentioned, it would be worthy to elaborate more on the coupled scalar field we presented when both torsion and non-metricity are allowed and direct couplings of the latter with the scalar field occur. Finally, it would be quite interesting to go beyond linear functions $F$  of the new dilation current term we considered. In this way   we will be able to investigate what exactly is the effect of this new  addition/extension as well as its phenomenology especially with regards to its energy momentum trace counterpart. 
	
	\section*{Acknowledgements}
	The work was  supported by the Ministry of Education and Science of the Republic of Kazakhstan, Grant  AP09058240.

		\bibliographystyle{unsrt}
	\bibliography{ref}

	\end{document}